\documentclass[epjST]{svjour}
\usepackage{graphics}
\usepackage{amsmath}
\usepackage{mathtools}
\usepackage{hyperref}
\usepackage[title]{appendix}
\begin{document}
\title{Demographic noise can promote abrupt transitions in ecological systems}
\author{Sabiha Majumder\inst{1,2}\fnmsep\thanks{\email{sabiha.physics@gmail.com}}, 
Ayan Das \inst{2}\fnmsep\thanks{\email{ayan.chromium@gmail.com}}, 
Appilineni Kushal \inst{1}\fnmsep\thanks{Current address: Department of Mathematics, University of California, Davis},
Sumithra Sankaran \inst{2}\fnmsep\thanks{Current address: Gesundheitsgeographie und Politik, ETH, Zurich}, 
\and 
Vishwesha Guttal\inst{2}\fnmsep\thanks{\email{guttal@iisc.ac.in}}}
\institute{Department of Physics, Indian Institute of Science, Bengaluru 560 012, India \and Centre for Ecological Sciences, Indian Institute of Science, Bengaluru 560 012, India}
\abstract{Strong positive feedback is considered a necessary condition to observe abrupt shifts of ecosystems. A few previous studies have shown that demographic noise -- arising from the probabilistic and discrete nature of birth and death processes in finite systems --  makes the transitions gradual or continuous. In this paper, we show that demographic noise may, in fact, promote abrupt transitions in systems that would otherwise show continuous transitions. We present our methods and results in a tutorial-like format. We begin with a simple spatially-explicit individual-based model with local births and deaths influenced by positive feedback processes. We then derive a stochastic differential equation that describes how local probabilistic rules scale to stochastic population dynamics. The infinite-size well-mixed limit of this SDE for our model is consistent with mean-field models of abrupt regime-shifts. Finally, we analytically show that as a consequence of demographic noise, finite-size systems can undergo abrupt shifts even with weak positive interactions. Numerical simulations of our spatially-explicit model confirm this prediction. Thus, we predict that small-sized populations and ecosystems may undergo abrupt collapse even when larger systems - with the same microscopic interactions - show a smooth response to environmental stress.} %

\maketitle
\section{Introduction}
\label{sec-intro}
Ecosystems can exhibit alternative stable states at similar environmental conditions~\cite{may1977nature,scheffer2001nature}. For example, in semi-arid ecosystems, vegetated state and bare state can co-exist at similar values of mean annual precipitation. Similarly, lakes can exist in turbid or clear states for the same amount of nutrient loading rates. Such systems with bistable (or multiple stable) states are prone to abruptly shift from one stable state to another when they cross a threshold value of the external driver. These systems also show hysteresis, i.e. the reverse transition occurs at a driver value different from the conditions that caused the initial transition~\cite{may1977nature}. Therefore, restoration is often difficult and sometimes even impossible. To explain these widely observed phenomena, strong positive feedback within ecosystems is often invoked as a necessary condition~\cite{taylor2005allee,xu2015local,kefi2016can,sankaran2019mee}. Here, we present a new insight on how demographic noise -- stochastic effect arising from discrete and probabilistic birth and death events in finite systems -- can exacerbate these effects even in systems with weak positive interactions. 

In ecological systems with positive feedback, organisms interact to enhance the local birth rates and/or reduce the local death rates. Many analytical and numerical simulation models incorporate  ecosystem specific positive interactions~\cite{taylor2005allee,sankaran2019mee,guichard2003amnat,dakos2011AmNat,kefi2007nature,von2010periodic,manor2008facilitation,scanlon2007positive,couteron2001periodic}. For example, in semi-arid ecosystems, plants increase local infiltration of surface water and reduce evaporation; thus, the chance of germination and growth of a plant nearby another plant is higher than that in a bare region. Such positive density dependence on the growth rates -- also called Allee effect -- is an important factor in maintaining multiple stable states and in driving abrupt transitions between these states~\cite{taylor2005allee,xu2015local,kefi2016can,sankaran2019mee}.  %

Besides interaction processes like positive feedback, stochasticity also plays an important role in determining ecosystem dynamics and stability. Broadly, stochasticity can be classified into two types: environmental (extrinsic) stochasticity arising from random fluctuations in the environmental drivers and demographic (intrinsic) stochasticity arising from probabilistic and discrete nature of birth and death of individuals. While it is known that environmental stochasticity can alter resilience and induce shifts between bistable ecological systems \cite{guttal2007ecomodel,sharma2015theoeco,chen2018amnat,meng2020interface,lucarini2019prl}, the effect of demographic stochasticity on the resilience of ecosystems has received less attention~\cite{meng2020interface,dennis2002allee,weissmann2014stochastic,martin2015PNAS,sarkar2020}. For example, demographic noise may smoothen out abrupt transitions~\cite{weissmann2014stochastic,martin2015PNAS}. In contrast, some studies make an opposite prediction too, i.e. demographic noise promotes abrupt transitions~\cite{dennis2002allee}. However, these studies use a phenomenological approach to incorporate noise in the models. In this approach, the direct relationship between the individual-level processes and the structure of noise at the population level are not rigorously established. Although demographic noise is now relatively well studied in a number of ecological contexts~\cite{mobilia2007jstat,black2012TREE,rogers2012epl,jhawar2019book,jhawar2020natphys,dobramysl2018jphysA}, the precise role of finite population size on the dynamics of abrupt shifts is not much explored. Furthermore, few studies disentangle the effects of spatial interactions from demographic stochasticity on stability and resilience of ecosystems (but see~\cite{dennis2002allee,realpe2013ecocomp,lande1998demographic,ovaskainen2010stochastic}). 

In this paper, we explore the effects of demographic stochasticity arising from finite population sizes on ecological transitions. The manuscript is organised as follows, with an effort to make it pedagogical and accessible to anyone who is familiar with basic non-linear dynamics and stochastic processes. In Section~\ref{sec-model}, we present a minimal, spatially-explicit individual-based model which incorporates local processes of birth and death and positive feedback interactions among individuals~\cite{sankaran2019mee,lubeck2006JStatPhys,sankaran2018ecoind}. We then present van Kampen's system-size expansion method in Section~\ref{sec-meso-sde} which yields a mesoscopic model, i.e. a stochastic differential equation governing the system dynamics while accounting for finite-size stochastic effects.  Using this method, we derive the mesoscopic description of our individual-based spatial model and relate microscopic interaction rules to the structure of demographic noise. In Section~\ref{sec-mft}, we show results of the model in the deterministic mean-field limit. In Section~\ref{sec-demo-noise}, using analytical calculations and numerical simulations, we show that demographic stochasticity promotes alternative stable states even with relatively weak positive interactions. %
Finally, we discuss various implications, including the counteracting effects of demographic noise and spatial interactions on ecosystem transitions.

\section{A minimal individual-based model for ecosystem transitions}
\label{sec-model}
\begin{figure}[ht]
\resizebox{1.0\columnwidth}{!}{%
 \includegraphics{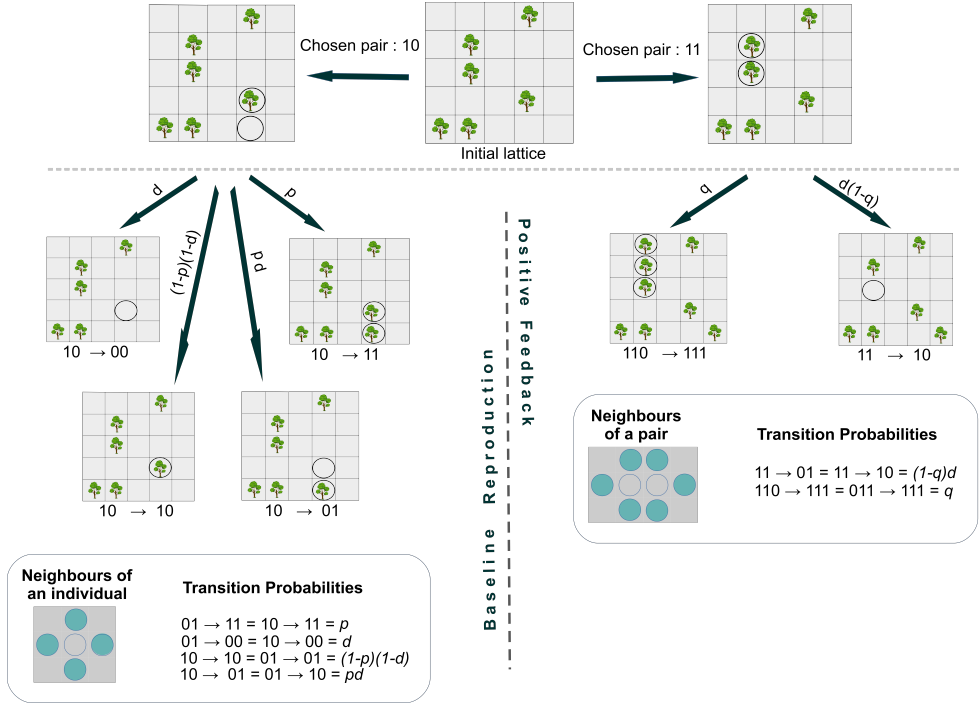} }
\caption{\textbf{Schematic of the model update rules}. Left panel shows the baseline reproduction process as implemented when the neighbouring site of a focal plant is empty. Right panel shows the positive feedback process which is implemented when the neighbouring site of a focal plant is occupied. (Schematic adapted with modification from Figure 1 in the Supporting Information for~\cite{sankaran2019mee})}
\label{fig-model-schematic}       %
\end{figure}

Our model is based on a statistical-physics inspired cellular-automaton model of tricritical directed percolation~\cite{lubeck2006JStatPhys} that has been adopted in ecological contexts~\cite{sankaran2019mee,sankaran2018ecoind,majumder2019ecology}. Here, the two dimensional space is divided into $L\times L = N$ discrete sites. Each site can take one of the two states: empty (denoted by 0) or occupied (by an organism; denoted by 1). We update these states at discrete time steps using probabilistic rules of birth and death, described below (See Fig \ref{fig-model-schematic} for a schematic of the update rules of the model).

{\it Baseline reproduction:} During each discrete time step, a site is selected randomly, henceforth called the focal site. If the focal site is empty, no update is done and another site is randomly selected. However, if the focal site is occupied by an organism, one of its four nearest neighbours is selected at random. This neighbour may be occupied or empty. If it is empty, we implement what we refer to as the baseline reproduction rule: the focal plant reproduces with a probability $p$ and establishes a new plant at the neighbouring empty site (resulting in a transition represented as 10 $\to$ 11). In addition, the focal plant may die with a probability $d$ (represented as 10 $\to$ 00). 

We interpret $p$ as the baseline birth probability of the plants (or more generally an organism) per unit time which can be thought of as a parameter representative of external factors like precipitation; we also refer to $p$ as the environmental driver and a reduction in $p$ as increased environmental stress. We interpret $d$ as the density-independent death rate; this is in contrast to the previous models where the death rate was assumed to be $1-p$~\cite{sankaran2019mee,lubeck2006JStatPhys,sankaran2018ecoind,majumder2019ecology}.  

{\it Positive feedback}: If the neighbouring site of the focal site, however, is occupied, we implement the  positive feedback rule: we randomly pick one of the six neighbours of the pair as the third site. If the third site is empty a plant germinates there with a probability $q$ (represented as 110 $\to$ 111) or the focal plant dies with a probability $d$ (represented as 11 $\to$ 01); the latter rule is implemented only if the third site is not germinated. The positive feedback (i.e. $q>0$) has two effects: It enhances the birth rate of the plants. In addition, there is reduction of death rate of the focal plant when there are occupied neighbors. 

Therefore, a discrete step update may involve (i) nothing if the focal site is empty, (ii) a basic reproduction rule if the chosen pair is 10, leading to one of the four outcomes shown in the left panel of Fig \ref{fig-model-schematic} or (iii) a positive feedback rule if the chosen pair is 11, leading to one of the two outcomes shown in the right panel of Fig \ref{fig-model-schematic}. One time unit completes after $L^2$ discrete updates such that all the cells are selected, on average, once for an update. %

We now make remarks on how this model relates to {\it contact process}, a well known spatial stochastic process studied in the context of infectious disease spread and population dynamics~\cite{da1999PRE,morris1995spread}. Unlike the contact process, we assume that reproduction by the focal plant and its death are not mutually exclusive. We model birth and death as independent processes and both can happen in the same discrete time step, corresponding to the update $10 \rightarrow 01$ which occurs with a probability $pd$. Likewise, there is a finite chance of neither of these events occurring, corresponding to the update $10 \rightarrow 10$ with a probability $(1-p)(1-d)$. Therefore, the update rules of our model do not converge to the discrete formulation of the contact process by setting $q=0$ and $d=1-p.$ 

However, we stress that $10 \rightarrow 10$ and $10 \rightarrow 01$ transitions do not affect analytical approximations like mean-field and pair approximations since these updates have no effect on the macroscopic density of plants on the landscape. See Section~\ref{sec-meso-sde} and Appendix A.  Hence, in the mean-field limit, our model with $q=0$ and $d=1-p$ is equivalent to the contact process. Likewise, these transitions do not affect the doublet densities either; therefore, the pair-approximation for the contact process and our model with $q=0$ and $d=1-p$ are identical.

\section{Coarse-grained population level model with demographic noise}
\label{sec-meso-sde}

Mean-field approximation of stochastic cellular-automaton models are constructed in the macroscopic limit ($N \to \infty$); thus, apart from spatial interactions, they fail to account for the stochasticity arising from finite system sizes \cite{mobilia2007jstat,mckane2005prl,biancalani2014PRL}. In this section, we first illustrate how the so called {\it system-size expansion method} can be used to obtain a mesoscopic dynamical description - which approximates the dynamics of a coarse-grained state variable such as population density and its dependence on the size of the system~\cite{jhawar2019book,biancalani2014PRL,van1992book}. We first derive the SDE using system size expansion method for a generic set of stochastic update rules which determine the value of a state variable $x$. We then apply the method to the model described in the previous section. %

\subsection{Deriving the mesoscopic model with system-size expansion}
\label{sec-system-size-expansion}

We would like to describe the dynamics of a scalar variable $\rho$, representing the proportion of occupied cells or density of the landscape. For the sake of generality, we use the notation $x$ for the state variable, since the subsequent steps outlined in this section are applicable to a system of reactions which can be described by a single dynamical state variable. We further assume that in a finite system of size $N$ the stochastic update rules are `birth' or `death' like events, which may change $x$ to $x+\frac{1}{N}$ or $x-\frac{1}{N}$. 

To begin with, we write the master equation to describe the temporal evolution of the probability distribution of $x$ at time $t$~\cite{van1992book,gardiner1985book}, denoted by $P(x,t)$ ,

\begin{equation}
\label{eq-masterequation}
\begin{split}
    \frac{\partial P(x,t)}{\partial t} &= \sum_{x' \ne x} [T(x|x')P(x',t) - T(x'|x)P(x,t)] \\
    & = T(x|x+\frac{1}{N}) P(x+\frac{1}{N},t) + T (x|x-\frac{1}{N}) P(x-\frac{1}{N},t)  \\
    &- T(x+\frac{1}{N}|x) P(x,t) - T (x-\frac{1}{N}|x) P(x,t) 
\end{split}
\end{equation}

\noindent where $T(x|x')$ is the transition rate from the state $x'$ to state $x$, which in turn depends on the stochastic update rules of the model.

Following~\cite{jhawar2019book,biancalani2014PRL}, to simplify the above master equation, we define the step operators $\epsilon^+$ and $\epsilon^-$ as 
\begin{equation*}
    \epsilon^+ h(x) = h(x+\frac{1}{N}) \quad \quad \quad \epsilon^- h(x) = h(x-\frac{1}{N})
\end{equation*}
\noindent In addition, for brevity, we denote the transition rates that correspond to birth and death as $T^+$ and $T^-$, respectively,
\begin{equation*}
    T^+(x)  =  T (x+\frac{1}{N}|x)\\ \quad \quad \quad
    T^-(x)  =  T(x-\frac{1}{N}|x)
\end{equation*}

Using these operators and the abbreviated notation for transition rates, we may rewrite the master equation (\ref{eq-masterequation}) as

\begin{equation}
    \label{eq-abb-master}
    \frac{\partial P(x,t)}{\partial t} = (\epsilon^- - 1) T^+(x) P(x,t) + (\epsilon^+ - 1) T^-(x) P(x,t)  
\end{equation}

Assuming large $N$, we approximate our step operators with a second order Taylor series:

\begin{equation}
    \label{eq-step-approx}
    \epsilon^{\pm} h(x)  \approx \left[1 \pm \frac{1}{N} \frac{\partial}{\partial x} + \frac{1}{2N^2} \frac{\partial^2}{\partial x ^2} \right]h(x)
\end{equation}

Substituting Eq (\ref{eq-step-approx}) in Eq (\ref{eq-abb-master}), distributing terms and rearranging, we have

\begin{equation*}
    \begin{split}
    \frac{\partial P(x,t)}{\partial t} & \approx - \frac{1}{N} \frac{\partial}{\partial x} \left[\left(T^+(x) - T^-(x)\right) P(x,t)  \right] + \frac{1}{2N^2} \frac{\partial^2}{\partial x ^2} \left[\left(T^+(x) + T^-(x)\right) P(x,t)  \right]
\end{split}
\end{equation*}

To ease the notation, we make the following substitutions
\begin{equation}
 f(x) = T^+(x) - T^-(x) \quad \quad \quad g(x) = \sqrt{T^+(x) + T^-(x)}\\
\label{eq-f-and-g}
\end{equation}

Next, we rescale time as $t=Nt'$ and drop the prime to obtain

\begin{equation}
\label{eq-FPE}
\frac{\partial P(x,t)}{\partial t} = - \frac{\partial}{\partial x} \left[f(x) P(x,t) \right] + \frac{1}{2N}  \frac{\partial^2}{\partial x ^2} \left[g^2(x)P(x,t) \right]
\end{equation}

This is called the Fokker-Plank equation (FPE). For a given stochastic process, FPE describes the time evolution of the state variable's probability distribution $P(x,t)$. If we wish to analyse individual trajectories, we must construct a stochastic differential equation (SDE) for the state variable. For a given stochastic process, the SDE and FPE are closely related~\cite{van1992book,gardiner1985book}. The {\it Ito sense} SDE corresponding to the the FPE (\ref{eq-FPE}) is given by~\cite{van1992book,gardiner1985book}
\begin{equation}
\label{eq-sde}
\frac{dx}{dt} = f(x) + \frac{1}{\sqrt{N}}g(x) \eta(t)
\end{equation}

\noindent where $\eta(t)$ is a Gaussian white noise with $\left< \eta(t) \right> = 0$ and $\left<\eta(t)\eta(t') \right> = \delta(t-t').$ Eq~(\ref{eq-sde}) together with Eq~(\ref{eq-f-and-g}) constitute the mesoscopic description of the microscopic stochastic model of our interest. We remark that the stochastic term in this description has two features: First, the noise is multiplicative, i.e. the strength of noise depends on the current value of the state variable $x$. Secondly the strength also depends, inversely, on the system size $N$. 

\subsection{Steady-states and bifurcation diagram of mesoscopic model} \label{sec-steady-bif}

To obtain the steady-state probability distribution function $P_s(x)$, we set $\frac{\partial P_s(x,t)}{\partial t}  = 0$ in Eq~(\ref{eq-FPE}). We then solve the resulting ordinary differential equation under the {\it no current} condition at the boundary. This is equivalent to assuming that $\partial_x P(x) \to 0$ as $x \to \infty$ and $P(x) \to 0$ as $x \to \infty$; the latter is a necessary condition for the normalisation of probability density function.  This yields 

\begin{equation}
P_s(x) = \frac{1}{Z} \exp(- 2 \, N \, U(x)) 
\end{equation}
\noindent where $U(x)$ is referred to as effective potential and is given by~\cite{horsthemke1984noise}
\begin{equation}
U(x) = - \int_{x_0}^{x} \frac{f(y) - \frac{1}{N} g(y) g'(y)}{g(y)^2} dy
\end{equation}
\noindent where $x_0$ is an arbitrary value in the physical domain of the state variable.

Bifurcation diagram is typically described for mean-field deterministic systems (but see ~\cite{guttal2007ecomodel}). Here, to obtain the bifurcation diagram of the mesoscopic description which contains the noise term, we identify the extrema of $P_s(x)$.  If the extrema correspond to a mode of $P_s$, it corresponds to a most likely state which can be considered equivalent to a stable state with noise. On the other hand, if the extrema is a minima of $P_s$, we argue that it corresponds to an unstable state in a noisy system. We also note the trivial solution $\rho^*=0$ of  Eq~(\ref{eq-FPE-model}) is an absorbing state and is also consistent with $\rho^*=0$ being a steady-state solution of the mesoscopic SDE (Eq~\ref{eq-sde_model}).  To plot the bifurcation diagram, we plot both the trivial solution and nonzero solutions (extrema of $P_s$) as a function of $p$, for different values of $N$ (Fig~\ref{fig-finite-analytical}). 

While the above procedure can in principle be done analytically, there may be no closed form solutions for $P_s(x)$. In such cases, one may use numerical integrator schemes and obtain the bifurcation diagram.

\subsection{Mesoscopic dynamics (SDE) of our model}
\label{sec-sde-pqd}

We did not use any specific form for the transition rates in the previous subsection to emphasize the generality of system-size expansion in deriving mesoscopic models. 

Now, we explicitly write transition rates for the density $\rho$ -- the proportion of occupied sites in the cellular-automaton landscape -- and obtain an SDE for its dynamics. In the limit of a well-mixed system, specifying the transition rates is a straightforward application of the law of mass action: the rate of reaction is proportional to the concentration of the reacting species. Applying this principle, we obtain

\begin{equation}\label{eq-transition1}
    T^+(\rho) =  T(\rho + \frac{1}{N}|\rho) = \rho(1-\rho)p + \rho^{2}(1-\rho)q 
\end{equation}

\begin{equation}\label{eq-transition2}
     T^-(\rho) = T(\rho - \frac{1}{N}|\rho) = \rho(1-\rho)d + \rho^{2}(1-q)d
\end{equation}

In Eq (\ref{eq-transition1}) we have used the fact that the density can change from $\rho$ to $\rho + \frac{1}{N}$ in two ways: $01 \rightarrow 11$ with a probability $p$ or $011 \rightarrow 111$ with a probability $q$.  Similarly, in Eq~(\ref{eq-transition2}) the density can change from $\rho$ to $\rho-\frac{1}{N}$ in two ways: $10 \rightarrow 00$ with a probability $d$ or $11 \rightarrow 01$ with a probability $(1-q)d$.  See Appendix~\ref{sec-app-mfa} for further detailed steps on well-mixed or mean-field approximation. 

We emphasise that reactions $01 \rightarrow 01$ and $01 \rightarrow 10$ have no effect on $\rho$, hence they do not appear in the master equation. 

Substituting Eq~(\ref{eq-transition1}) and Eq~(\ref{eq-transition2}) in Eq~(\ref{eq-sde}), we obtain the mesoscopic SDE for our model, interpreted in an {\it Ito sense}, 

\begin{eqnarray}
\label{eq-sde_model}
\frac{d\rho}{dt} & = & (p-d)\rho - (p-q-qd)\rho^{2} - q\rho^{3} \nonumber \\ & + & \frac{1}{\sqrt{N}} \sqrt{(p+d)\rho -(p-q+qd)\rho^{2} -q\rho^{3}} \, \eta(t)
\end{eqnarray}

\noindent where $\eta(t)$ is a Gaussian white noise with zero mean and a unit variance. 

The mesoscopic SDE (\ref{eq-sde_model}) has two terms: The first term is the deterministic part and is same as the mean-field approximation (see Appendix~\ref{sec-app-mfa}). The second term is the stochastic part,  where the strength of noise depends on the state of the system ($\rho$) and is inversely related to the system size $N$. Therefore, via Eq~(\ref{eq-sde_model}) we are able to capture the demographic noise -- which arises from stochasticity of birth and death in finite systems.  

In the next two sections, we analyse two scenarios of the above mesoscopic description. First we analyse the steady-states in the macroscopic limit $N \to \infty$ -- which is same as the mean-field approximation -- and then we analyse the effect of demographic noise for finite $N$.  

\section{Mean-field approximation shows that positive feedback promotes alternative stable states in infinitely large systems}
\label{sec-mft}

\begin{figure}[t]
\resizebox{1.0\columnwidth}{!}{%
 \includegraphics{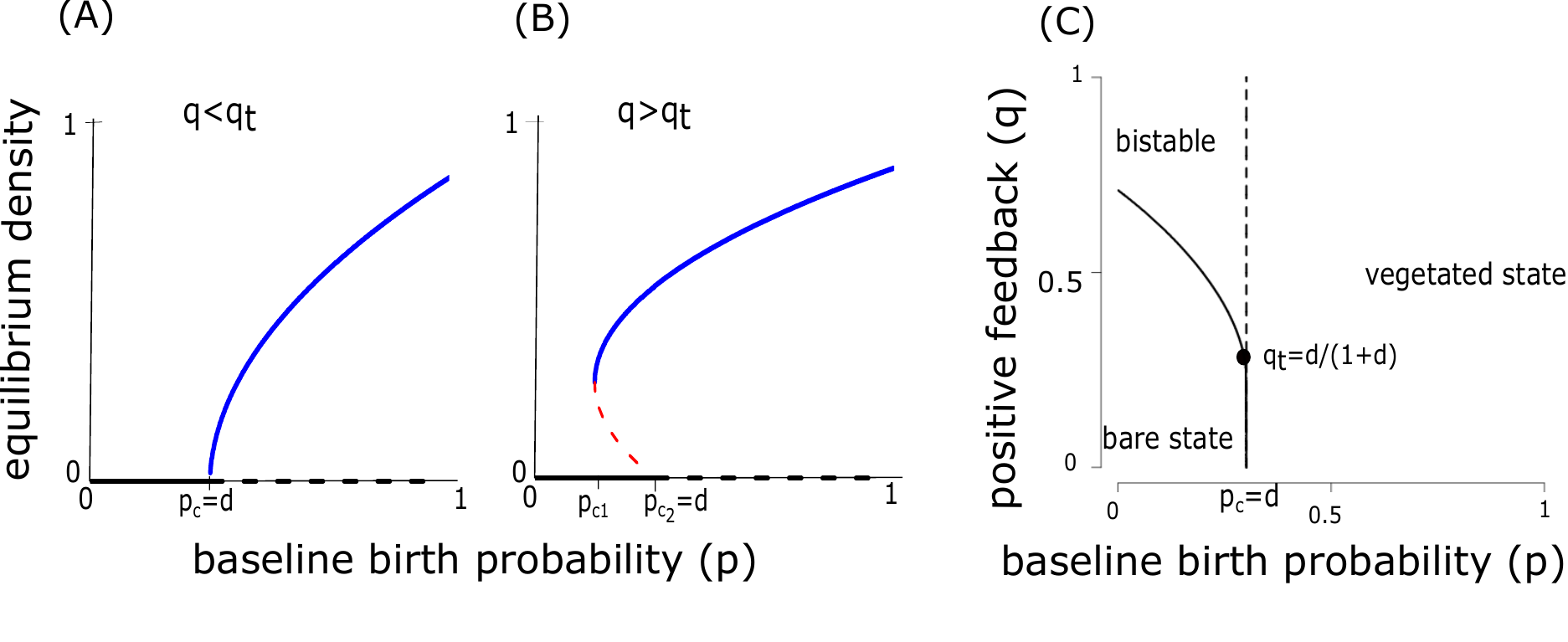} }
\caption{\textbf{Mean-field model shows continuous transition (A) or discontinuous transition (B), depending on the value of the positive feedback parameter $q$}. Black, blue and red curves show the three equilibrium densities. Solid lines are the stable states and dotted lines are the unstable states. For low values of positive feedback ($q<q_t$), the system shifts from a vegetated state to a bare state ($\rho=0$) continuously at $p_c=d$. For strong positive feedback ($q>q_t$), two stable states exist for a range of parameter value: $p_{c2}-p_{c1}$. When $p$ is reduced below a threshold ($p_{c_1} = d-\frac{(p_{c_1}-q-qd)^2}{4q}$, hence lower than $d$), the system shifts from a vegetated state to bare state abruptly. The reverse transition occurs at the $p_{c_2}=d$ \textbf{(C)} shows the stability diagram of the mean-field model for all values of $p$ and $q$. Solid line shows the critical thresholds ($p_{c1}$) below which only bare state is stable. Black dot represents the tri-critical point at which the nature of transition changes from continuous to discontinuous. The region between the solid line and the dotted line has two states. Here, $d$ is kept fixed as $0.3$.}
\label{fig-mean-field}       %
\end{figure}

Considering the macroscopic limit, i.e.  $N \rightarrow \infty$, the stochastic differential equation converges to an ordinary (deterministic) differential equation given by

\begin{equation}
\frac{d\rho}{dt} = (p-d)\rho - (p-q-qd)\rho^2 - q\rho^3
\label{eq-meanfield}
\end{equation}

We obtain the equilibria of the mean-field model by setting $d\rho/dt = 0$ and find their stability; we refer the reader to Appendix A for detailed steps. One of the equilibria is a bare state $\rho*=0$. Based on the linear stability analysis~\cite{strogatz1994nonlinear}, we find that the bare state is stable when $p<d$ and is unstable for $p>d$. In other words, when the baseline birth rate $p$ exceeds a critical threshold of $p_c=d$, the system in bare state undergoes a transition to a vegetated state ($\rho^*>0$).  %

The nature of the transition as a function of $p$, however, depends on the value of the positive feedback parameter $q$.  Based on linear stability analysis, we find that the transition from a bare state to a vegetated state, as a function of $p$, is a continuous transition for $0 \le q<\frac{d}{1+d}$, i.e. low values of $q$. In other words, for ecosystems with weak positive interactions, the transition from a vegetated to a bare state is smooth and gradual function of the environmental driver (Fig~\ref{fig-mean-field} A). 

For systems with strong positive feedback, specifically $q >q_t = \frac{d}{1+d}$,  the transition between bare state and the vegetated state is a discontinuous function of $p$ (Fig~\ref{fig-mean-field}B). Here, for a range of driver values $p_{c_1}<p<p_{c_2}$, where $p_{c_1}=d-\frac{(p_{c_1}-q-qd)^2}{4q}$ and $p_{c_2}=d$, we find three equilibria; a stable bare state (solid black line in Fig~\ref{fig-mean-field} B), an unstable intermediate vegetated state (dotted red line in Fig~\ref{fig-mean-field} B) and a stable large vegetated state (solid blue line in Fig~\ref{fig-mean-field} B).  The value of parameter $p$ at which the stable and unstable vegetated states meet and cease to exist, denoted $p_{c1}$, is referred to as a saddle-node bifurcation in the non-linear dynamics literature. 

If a system is in the vegetated state and the driver values reduces below  the critical threshold of $p_{c1}$ ($<d$), the system undergoes an abrupt transition from a vegetated state to the bare state. Likewise, when the driver increases above the threshold $p_{c2}=d$, a system in bare state undergoes an abrupt transition to a vegetated state. These transitions are also referred to as abrupt regime shifts, catastrophic shifts, tipping points or critical transitions~\cite{scheffer2001nature,scheffer2009nature}. We also note that the system exhibits hysteresis; i.e. the two abrupt transitions occur at different threshold values of $p$. Furthermore, depending on the initial condition, the system may reach either the bare state or the vegetated state. 

The threshold value of positive feedback at which the type of transition changes from smooth to abrupt ($q=\frac{d}{1+d}$) is called the tri-critical point. The region of bistability, i.e. the range $p_{c_1}$ to $p_{c_2}$,  increases with the strength of positive feedback $q$. 

Fig \ref{fig-mean-field} C summarises the mean-field predictions of our spatially-explicit stochastic model, as a function of environmental driver (baseline birthrate $p$) and positive feedback parameter ($q$). It shows that vegetated systems with strong positive feedback can survive in harsher environmental conditions (i.e. for lower values of $p$ than the critical threshold of continuous transitions; note that $p_{c1}<d$). However, systems with strong positive feedback are also more prone to collapse as a function of changing environmental conditions. In other words, within the mean-field approximation (i.e. in well-mixed, infinitely large systems), large positive feedback among individuals promotes bistability in ecosystems. 

The mean-field approximation results and interpretations of our model are consistent with the literature on spatial ecosystem models with positive feedback~\cite{kefi2016can,guichard2003amnat,kefi2007TPB,kefi2014PlosOne}, which often incorporate models of ecosystems with many more parameters (but see~\cite{sankaran2019mee,majumder2019ecology}). We now move onto studying the role of demographic noise.

\section{Demographic noise can induce alternative stable states even with weak positive feedback}
\label{sec-demo-noise}

To investigate the effects of demographic noise induced by finite system-sizes, we first calculate the steady-state probability density function of our mesoscopic model (Eq~(\ref{eq-sde_model})) with $q=0$, i.e in the weak positive feedback limit. To do so, we revert to the corresponding Fokker-Planck equation 
\begin{equation}
\label{eq-FPE-model}
\frac{\partial P(\rho,t)}{\partial t} = - \frac{\partial}{\partial \rho} \left[\left((p-d)\rho-p\rho^2\right) P(\rho) \right] + \frac{1}{2N}  \frac{\partial^2}{\partial \rho ^2} \left[\left((p+d)\rho-p\rho^2\right) P(\rho) \right] 
\end{equation}

\noindent where the strength of demographic noise is proportional to $\frac{1}{\sqrt{N}}$.  Using the method described in Section~\ref{sec-steady-bif}, we compute the steady-state $P_s(\rho)$ and the bifurcation diagram of the mesoscopic model. 

\begin{figure}[t]
\resizebox{1.0\columnwidth}{!}{%
 \includegraphics{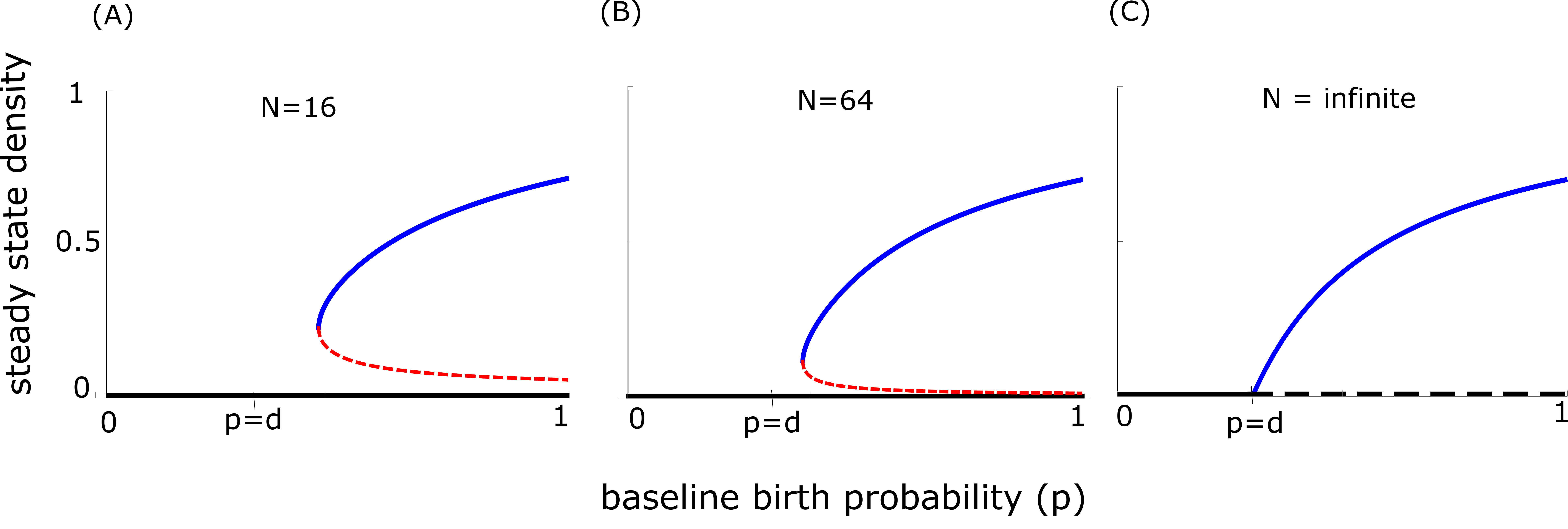} }
\caption{\textbf{Bifurcation diagrams constructed by finding the extrema of the steady-state probability distribution for the FPE of our model shows that finite systems undergo abrupt shifts.} Solid lines are the stable states and dotted lines are the unstable states in the stochastic model. The steady-states are derived from the FPE with $q=0$. (A) and (B) show that finite systems have alternative stable states and they show abrupt transition. The bifurcation diagram converges to the mean-field bifurcation diagram as $N \rightarrow \infty$ as shown in (C). Here, $d=0.3$ and $q=0$.}
\label{fig-finite-analytical}       %
\end{figure}

We find that for finite systems, the transition from vegetated to bare state is discontinuous (Fig~\ref{fig-finite-analytical} A and B). Furthermore, the transition occurs at a threshold value of the driver $p$ which is larger than the mean-field threshold (i.e $p_c>d$), implying that the collapse to bare state occurs in conditions in which larger systems would survive in the vegetated state. Therefore, we predict that small systems may exhibit alternative stable states and abrupt collapse due to demographic noise. As shown in section~\ref{sec-mft}, in the macroscopic limit $N \to \infty$, the SDE converges to mean-field model and the system shows continuous transition from vegetated state to bare state at $p_c=d$ (Fig ~\ref{fig-finite-analytical} C).  %

\begin{figure}[t]
\resizebox{1.0\columnwidth}{!}{%
 \includegraphics{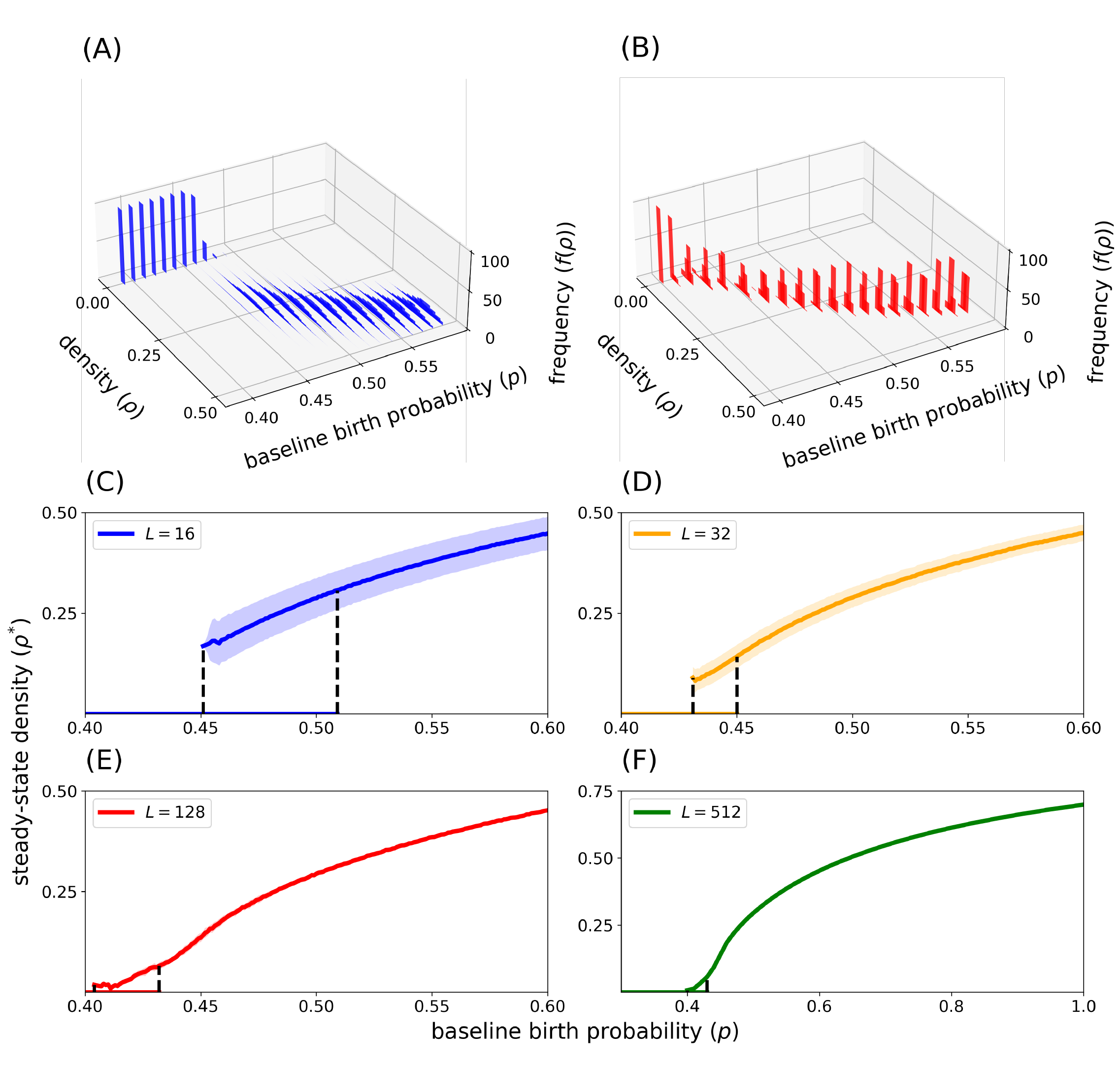} }
\caption{\textbf{Numerical simulations of the spatially-explicit individual-based model confirm bistability for finite-size systems.} \textbf{(A)} Region between $p=0.45$ to $p=0.5$ shows bi-modality. Further, a range of densities with zero-probability of occurrence is evident between the non-zero mode and the mode at zero. \textbf{(B)} \textbf{$L=128.$}  There is no sign of bi-modality and the transition from $\rho^{*}=0$ to $\rho^{*}>0$ is much smoother. \textbf{(C), (D), (E)} and \textbf{(F)} show ``bifurcation diagrams" constructed from numerical simulations for different system sizes in two dimensions.}  
\label{fig-num-sim}       %
\end{figure}

To verify if these predictions hold true in spatially-explicit systems, we conducted extensive numerical simulations of our spatially-explicit individual-based model using the stochastic update rules described in Section~\ref{sec-model}. We display the results in Fig.~\ref{fig-num-sim}. For all parameter values (except for $p>0.65$ in $L=512$, where we simulated 10000 time steps to reach steady-state), we run the simulation for 1 million time steps so that the system reaches a steady-state. We chose four system sizes: $16 \times 16, 32 \times 32$, $128 \times 128$ and $512 \times 512$ for the purpose of illustration in Figure~\ref{fig-num-sim}. For the system size $16 \times 16$, for each value of  $p$ we construct a normalized steady-state frequency distribution of density, denoted $f(\rho)$, based on 10,000 independent realizations (Fig~\ref{fig-num-sim} A). For $128 \times 128$, due to computational constraints, we calculate $f(\rho)$ based on 100 realisations only (Fig~\ref{fig-num-sim} B). We then calculate the `bifurcation diagram' for $L = 16$ (Fig~\ref{fig-num-sim} C), $32$ (Fig~\ref{fig-num-sim} D), $128$ (Fig~\ref{fig-num-sim} E), $512$ (Fig~\ref{fig-num-sim} F) based respectively on 10000, 2000, 100 and 100 realisations as follows: for each $p$, the upper arm was constructed by culling the realizations that fell into the absorbing phase ($\rho^{*}=0$) and calculating the mean of the densities of the remaining realisations. The lower arm $(\rho^{*}=0)$ was plotted if at least one realization with $\rho^{*}=0$ was reported. This method gives a well defined bi-stable region for each $L.$ 

Indeed, we confirm the qualitative predictions: small systems which have higher demographic noise exhibit bimodal frequency distribution of steady-state density~(\ref{fig-num-sim}A, C, D) and thus may undergo abrupt transitions from a vegetated state to a bare state. Furthermore, we  confirm the analytical prediction that smaller system sizes undergo abrupt transition from the vegetated to bare state at larger values of the driver $p$ in comparison to the macroscopic limit (Fig~\ref{fig-num-sim}C-E). We also find that the region of bistability reduces with increasing system size (Fig~\ref{fig-num-sim}C-D). Finally, larger systems are likely to undergo continuous transitions (\ref{fig-num-sim}B, F).

\section{Discussion}
\label{sec-disc}

In this study, we used a spatially-explicit individual-based  model to investigate the role of demographic stochasticity in ecological transitions.  Using analytical methods and numerical simulations, we showed that  demographic noise in finite systems can cause bistability and abrupt transitions even with weak positive feedback mechanisms. In other words, we predict that smaller ecosystems can undergo abrupt regime shifts even when larger ecosystems will survive under the same environmental conditions.

Our conclusions are broadly consistent with the results in the ecology literature, which has extensively looked at the role of stochasticity in population extinctions. Specifically, these studies conclude that demographic stochasticity can induce Allee-like effects, in which populations smaller than a threshold size are likely to go to extinction~\cite{dennis2002allee,lande1998demographic,ovaskainen2010stochastic}. However, these studies take phenomenological mean-field approaches, i.e. they construct the deterministic and stochastic terms of the dynamics using phenomenological heuristic. In contrast,  we begin with a spatially-explicit individual-based model with stochastic update rules. We then derive, using the method of system size expansion, how these microscopic update rules scale to mesoscopic dynamics of the population. This allowed us to explicitly capture the effect of finite system size in the stochastic term. Our derivations show that the stochastic term is a multiplicative noise term, depending on both the state variable density ($\rho$) and the system size ($N$); this is of the form $\frac{1}{\sqrt{N}}\sqrt{\rho + \mathcal{O}(\rho^2)}$. However, most previous studies include the density effects via a $\sqrt{\rho}$ term alone  \cite{weissmann2014stochastic,butler2009PNAS,dOdorico2005PNAS,mankin2002PRE,mankin2004PRE} and do not consider the effect of system size. Therefore, we argue our bottom-up approach helps in calculating the precise structure of demographic noise from the individual-based rules. 

Some recent studies of ecosystem transitions use phenomenological spatial models with demographic stochasticity \cite{weissmann2014stochastic,martin2015PNAS,butler2009PNAS}. These models analyse the combined effects of spatial interactions (via a diffusion term) and the demographic noise (assumed to be proportional to $\sqrt{\rho}$). By including both spatial interactions and demographic noise in the models, they find that demographic noise smoothens ecosystem transitions. However, these studies do not disentangle the effect of spatial interactions from the effect of demographic noise.  Using our approach, by ignoring the spatial fluctuations in our approximations, we can disentangle the effect of demographic noise and spatial interactions. We conjecture that while demographic noise makes the system prone to abrupt transitions, spatial interactions can smoothen these transitions (as supported by pair approximation calculations in Appendix~\ref{sec-app-pair}).

Another form of stochasticity, environmental noise, is also very important for determining ecosystem dynamics. Many studies have shown that large environmental noise can induce abrupt transitions in ecosystems even before critical thresholds \cite{guttal2007ecomodel,chen2018amnat,siteur2016oikos,ashwin2012tipping}. However, in some cases~\cite{dOdorico2005PNAS} environmental fluctuations can change the stability landscape and induce resilience. The combined effect of demographic noise and environmental noise needs to be carefully investigated to obtain insights for the management of ecosystems.

In conclusion, we show that even with limited positive feedback, finite systems can undergo discontinuous transitions. These findings can help us understand mechanisms of abrupt transitions in ecosystems. Ecological systems are finite in extent and are getting fragmented because of human activities. We predict that fragmented ecosystems are more sensitive to changes in external conditions than larger systems. A plethora of recent studies have investigated the possibility of forecasting abrupt transitions, using the idea of early warning signals~\cite{sharma2015theoeco,sarkar2020,scheffer2009nature,dakos2010TheorEco,guttal2008EcoLet,guttal2009TheorEco,carpenter2006EcoLet,carpenter2010ecosphere,dai2012science,eby2017GEB,barbier2006self,guttal2016PlosOne,lenton2011early,burthe2016early}. This begs an obvious question - are there early warning signals of transitions in systems with demographic noise? We suggest that researchers may investigate this and other related questions in the future.

\section{Acknowledgements}
VG acknowledges support from DBT-IISc partnership program and DST-FIST. SM, AD and SS acknowledge scholarship support from MHRD.  

\section{Codes}

Codes and data available at \url{https://github.com/tee-lab/demographic_noise}

\begin{appendices}
\section{Mean-field Approximation}\label{sec-app-mfa}

To write the mean-field equation for the dynamics of the model, we assume infinite size and no spatial structure in the ecosystem, meaning each site in the system is equally likely to be occupied. The probability of any site being occupied is same as the global occupancy. Therefore, the transition rates in this model are as follows. 
Transition rate for a site to change from 1 to 0 :  
$$ \omega(1 \rightarrow 0) = d (1-\rho) + (1-q)d\rho $$
(Here d is the death rate, not the symbol for a differential).
Similarly, transition rate from 0 to 1: 
$$ \omega(0 \rightarrow 1) = p \rho + q\rho^2 $$
Probability of finding a site in occupied(1) state : $P(1)=\rho$ and probability of finding a site in state in empty(0) site: $P(0)=(1-\rho)$

The master equation can then be written as: 

$$\frac{dP(1)}{dt} = \omega(0 \rightarrow 1)P(0) - \omega(1 \rightarrow 0) P(1)$$

Substituting the above transition rates,

\begin{equation}
\frac{d\rho}{dt}= f(\rho) = (p-d)\rho - (p-q-qd)\rho^2 - q\rho^3
\label{meanfield}
\end{equation}

For simplicity, the above equation can be written in terms of $a$, $b$ and $c$ as:
\begin{equation}
f(\rho) = a\rho - b\rho^2 - c\rho^3
\label{simplemeanfield}
\end{equation}

where $a=p-d$, $b=p-q-qd$ and $c=q$. At equilibrium, $\frac{d\rho}{dt}=0$. This gives the following equilibria:

$$\rho\text{*}=0 ,\text{  } \rho\text{*} = -\frac{b}{2c}+\sqrt[]{\frac{b^2}{4c^2} + \frac{a}{c}}\text{  and  }  \rho\text{*} = -\frac{b}{2c}-\sqrt[]{\frac{b^2}{4c^2} + \frac{a}{c}} $$
We call these solutions as $\rho_0$ , $\rho_A$ and $\rho_B$ respectively.
For the above equilibria to be stable, $f'(\rho\text{*})<0$.

From (\ref{simplemeanfield}), we get:
\begin{equation}
f'(\rho) = a\rho - b \rho^{2} -c \rho^{3}
\end{equation}

$$f'(\rho)|_{\rho_0} = a $$
$$f'(\rho)|_{\rho_A} = -2a - \frac{b^2}{2c} + b \sqrt[]{\frac{a}{c}+ \frac{b^2}{4c^2}} $$
$$f'(\rho)|_{\rho_B} = -2a - \frac{b^2}{2c} - b \sqrt[]{\frac{a}{c}+ \frac{b^2}{4c^2}} $$

In this analyses, we consider $c>0$ because $c=q$ in our model which is a probability. From the above three equilibrium densities, only real and positive solutions are realistic. Therefore, we will reject the negative solutions. For the non-zero solutions to be real, $a>-\frac{b^2}{4c}$

\subsection*{Case 1: $b>0$}
In this parameter region, given $a>-\frac{b^2}{4c}$, $\rho_B$ is always negative. Therefore, the mean-field equation has only two solutions $\rho_0$, which is stable for $a<0$ and $\rho_A$, which is stable for $a>0$. At $a=0$, $\rho_A = 0$ and it increases monotonically after that. The mean-field system undergoes a continuous transition (also called second-order transition or trans-critical bifurcation) at a critical point $a=0$ for all  values of $b>0$.

\subsection*{Case 2: $b<0$}
In this parameter regime, $\rho_0$ is stable for $a<0$ and unstable for $a>0$. $\rho_B$ is real and positive only when $\frac{-b^2}{4c} < a < 0$ and it is unstable in this regime. $\rho_A$ is stable for $ a \geq \frac{-b^2}{4c}$. At $ a = \frac{-b^2}{4c}$, $\rho_A = \frac{|b|}{2c}$. The mean-field system shows a saddle-node bifurcation at the point $(a,\rho^*) = (\frac{-b^2}{4c}, \frac{|b|}{2c})$. The transition from an active phase (vegetated state) to absorbing phase (bare state) is discontinuous.  

In the mean-field approximation, our model undergoes a continuous transition when $b>0$ and a discontinuous transition when $b<0$. The system has a tri-critical point at $a=0, b=0$ where the nature of transition changes from continuous to discontinuous. Now, translating it back to our system (Eq~(\ref{meanfield})) with parameters $p,q$ and $d$,the tri-critical point occurs at $p_t=d$ and $q_t=\frac{d}{1+d}$. For $q<q_t$, continuous phase transition occurs at $p_c=d$. However, the critical point ($p_{c_1}$) decreases as a function of $q$ when $q>q_t$. Therefore, in discontinuous regime, the vegetated state can sustain harsher conditions represented by low values of $p$ when strength of positive feedback among plants is high. However, when the external conditions pass a threshold, the system collapses abruptly to the bare state.

\section{Pair Approximation}\label{sec-app-pair}

In the mean-field approximation, we assumed that each site on the lattice has equal probability of being occupied by a plant and that there are no spatial fluctuations in the system. We now incorporate the local spatial effects in the master equation. We assume that the probability of occupancy of a site depends on the state of its nearest neighbours. Therefore, we introduce conditional probability $q_{i|j}$ defined as probability that a site is in state $i$ given its nearest neighbour is in state $j$. In this approximation we have singlet density $\rho_1$ (same as mean-field approximation) and an additional doublet density ($\rho_{ij}$) defined as the probability of a site being in state $i$ and its neighbour being in state $j$. The pair $ij$ has the following properties in this approximation. 

\begin{equation}
    \rho_{ij} = \rho_{ji}   \\ \quad \quad \quad \quad
    \rho_{ij} = \rho_i~q_{j|i}   \\ \quad \quad \quad \quad
    q_{j|i} = 1- q_{i|i}
    \label{eq-pair-prop}
\end{equation}

Note that $i$ and $j$ can be 0 or 1. For singlet density ($\rho_1$), the master equation can be written as: 

\begin{equation}
    \frac{d\rho_1}{dt} = \omega(0 \rightarrow 1)\rho_0 - \omega(1 \rightarrow 0) \rho_1
    \label{eq-pair-singlet-master}
\end{equation}

The transition rates $\omega(0 \rightarrow 1)$ and $\omega(1 \rightarrow 0)$ are defined as follows:

\begin{equation}
    \omega(0 \rightarrow 1)= p~q_{1|0} + q~q_{1|0}~q_{1|1}  \\ \quad \quad \text{and} \quad \quad
    \omega(1 \rightarrow 0) = d~q_{0|1} - (1-q)~d~q_{1|1}
    \label{eq-pair-trans-prob}
\end{equation}

Here $p$ is the baseline birth probability, $q$ is the positive feedback parameter and $d$ is the death probability. Substituting Eq (\ref{eq-pair-trans-prob}) in Eq (\ref{eq-pair-singlet-master}) and using the fact that $\rho_{ij} = \rho_{ji}$ and $\rho_{ij} = \rho_i~q_{j|i}$ (Eq~(\ref{eq-pair-prop})) we obtain the following form for the master equation

\begin{equation}
\frac{d\rho_1}{dt} = \left\lbrace p~q_{0|1} + q~q_{1|1}~q_{0|1} - d~q_{0|1} - (1-q)~d~q_{1|1} \right\rbrace \rho_1
\label{eq-pair-master-long-singlet}
\end{equation}

Similarly, for the doublet density ($\rho_{11}$), the master equation can be written as

\begin{equation}
\begin{split}
\frac{d\rho_{11}}{dt} &= \omega(10 \rightarrow 11)\rho_{10} + \omega(01 \rightarrow 11)\rho_{01} - \omega(11 \rightarrow 01)\rho_{11} - \omega(11 \rightarrow 10)\rho_{11}\\
&=2 \lbrace \omega(10 \rightarrow 11)\rho_{10}\rbrace - 2 \lbrace \omega(11 \rightarrow 10)\rho_{11}\rbrace
\end{split}
\label{eq-pair-doublet-master}
\end{equation}

The transition rate $\omega(10 \rightarrow 11)$ is defined as

\begin{equation}
    \omega(10 \rightarrow 11) = \frac{1}{z}~p + \frac{z-1}{z}~p~q_{1|0} + \frac{1}{z}~q~q_{1|1} + \frac{z-1}{z}~q~q_{1|0}~q_{1|1}
    \label{eq-pair-10-11}
\end{equation}

Here $z$ is the number of nearest neighbours. The first term represents the rate of the event in which the $1$ in the pair $10$ reproduces at $0$ with probability $p$. The second term represents the rate of the event in which a neighbour of $0$ in the pair $10$ other than the $1$ in the pair is also $1$ (we call such neighbours as \textit{non-pair neighbours}; this occurs with probability $q_{1|0}$) \textit{and} it reproduces at $0$ with probability $p.$ The next two terms represent the growth rate due to positive feedback. The third term represents the event in which a non-pair neighbour of the $1$ in $10$ is also occupied (this occurs with probability $q_{1|1}$) \textit{and} a birth occurs with enhanced probability $q$ at $0$ of the pair. The last term represents the event in which a non-pair neighbour of the $0$ in $10$ is occupied (this occurs with probability $q_{1|0}$) \textit{and} its next neighbour is also occupied (this occurs with probability $q_{1|1}$) \textit{and} a birth occurs at the $0$ of the pair with enhanced probability $q$. 

Similarly, we define the transition rate $\omega(11 \rightarrow 10)$ as 

\begin{equation}
    \omega(11 \rightarrow 10) = \frac{1}{z}(1-q)~d + \frac{z-1}{z}(1-q)~d~q_{1|1} + \frac{z-1}{z}d~q_{0|1} 
    \label{eq-pair-11-10}
\end{equation}

The first term represents the rate of the event in which a $1$ in the pair $11$ dies with a diminished probability $(1-q)d$ due to the facilitative interaction with the other $1$ of the pair. The second term represents the rate of the event in which a non-pair neighbour of $1$ in the pair $11$ (which we call \textit{focal} to distinguish it from the other $1$ of the pair) is occupied (this occurs with probability $q_{1|1}$) \textit{and} the \textit{focal} $1$ dies with a diminished probability $(1-q)d$ due to the facilitative interaction with this non-pair neighbour. The third term represents the rate of the event in which a non-pair neighbour of the \textit{focal} $1$, is $0$ (this occurs with probability $q_{0|1}$) \textit{and} the focal $1$ dies with probability $d.$

Substituting Eq (\ref{eq-pair-10-11}) and Eq (\ref{eq-pair-11-10}) in Eq (\ref{eq-pair-doublet-master}), the master equation for the doublet density takes the following form

\begin{equation}
\begin{split}
\frac{d\rho_{11}}{dt} &= 2 \left\lbrace \textmd{ } \lbrace \frac{1}{z} p  +\frac{z-1}{z} p~q_{1|0} + \frac{1}{z} q~q_{1|1} + \frac{z-1}{z} q~q_{1|0}~q_{1|1} \rbrace  \textmd{ } \rho_{10} \right\rbrace \\
  &-2\left\lbrace \lbrace \frac{1}{z}(1-q)~d + \frac{z-1}{z}(1-q)~d~q_{1|1} + \frac{z-1}{z}d~q_{0|1}\rbrace \textmd{ } \rho_{11} \right\rbrace 
\end{split}
\label{eq-pair-master-long-doublet}
\end{equation}
 We are interested in two dimensional landscapes modelled by a square lattice. We use a von-Newmann neighbourhood $(z=4)$ in all our analyses. Note that using the properties in Eq (\ref{eq-pair-prop}), we may write $\rho_{10}=(q_{0|1}/q_{1|1})\rho_{11}.$ Further, we may write $q_{1|1}=1-q_{0|1}.$ Substituting these in Eq (\ref{eq-pair-master-long-singlet}) and Eq (\ref{eq-pair-master-long-doublet}) we obtain

\begin{equation*}
    \frac{d\rho_1}{dt} = M_1 \rho_1 \\ \quad \quad \quad 
    \frac{d\rho_{11}}{dt} = M_{11} \rho_{11} 
\end{equation*}

$M_{1}$ and $M_{11}$ are defined as follows

\begin{equation}
M_1 = p~q_{0|1} + q~(1-q_{0|1})~q_{0|1} - d~q_{0|1} - (1-q)d~(1-q_{0|1}) 
\label{eq-pair-M1}
\end{equation}

\begin{equation}
\begin{split}
M_{11} &=  2 \left(~\frac{1}{4}~p~\frac{q_{0|1}}{(1-q_{0|1})} + \frac{3}{4} p~q_{1|0}~\frac{q_{0|1}}{(1-q_{0|1})} + \frac{1}{4} q~q_{0|1} +\frac{3}{4} q~q_{1|0}~q_{0|1} \right)\\
&- 2 \left(~\frac{1}{4}(1-q)~d + \frac{3}{4} (1-q)~d~(1-q_{0|1})+ \frac{3}{4} q_{0|1}~d ~\right)
\end{split}
\label{eq-pair-M11}
\end{equation}

$\rho_1 = 0$ and $\rho_{11}=0$ give the trivial equilibria. In addition,  $M_1 = 0$ and $M_{11}= 0$ provide other equilibria of the system. To ease the notation, we substitute $q_{0|1}={\bf a}$ in Eq (\ref{eq-pair-M1}), set $M_{1}=0$ and simplify to obtain a quadratic equation in $\bf a$  

\begin{equation}
q {\bf a^2} - \left[ p + q(1-d) \right] {\bf a} + d(1-q) = 0 
\label{eq-pair-a}
\end{equation}

Assuming $q \neq 0$, the two solutions ${\bf a_+} $ and ${\bf a_-}$ are:
\begin{equation}
{\bf a_+} =  \frac{p + q(1-d) + \sqrt{ \left[p+q(1-d)\right]^2 - 4q(1-q)d }}{2q}
\label{eq-pair-solutiona+}
\end{equation}

\begin{equation}
{\bf a_-} =  \frac{p + q(1-d) - \sqrt{ \left[p+q(1-d)\right]^2 - 4q(1-q)d }}{2q}
\label{eq-pair-solutiona-}
\end{equation}

Now, we substitute $q_{0|1}={\bf a}$ and $q_{1|0} = {\bf b}$ in Eq (\ref{eq-pair-M11}), set $M_{11}=0$ and simplify to obtain a solution for $\bf b$ assuming $3 \mathbf{a} [p + q(1- \mathbf{a})] \neq 0$

\begin{equation}
    \mathbf{b} = \frac{ (1-\mathbf{a})\lbrace{(1-q)d\left[1+3(1-\mathbf{a})\right] + \mathbf{a}\left(3d-q\right) \rbrace} - p\mathbf{a}}{3 \mathbf{a} [p + q(1- \mathbf{a})]}
    \label{eq-pair-solutionb}
\end{equation}

Density ($\rho_1$) can be calculated from  $q_{0|1}$ and  $q_{1|0}$ as following 
$$ \rho_1 = \frac{\rho_{10}}{q_{0|1}} = \frac{\rho_{01}}{q_{0|1}} = \frac{\rho_0~q_{1|0}}{q_{0|1}}  $$
$$  \rho_1~q_{0|1} = (1-\rho_1)~q_{1|0} $$
\begin{equation}
\rho_1  = \frac{q_{1|0}}{q_{0|1}+q_{1|0}} = \frac{{\bf b}}{{\bf a}+\bf{b}}  
\label{eq-pair-rho-from-ab}
\end{equation}

Therefore, the steady-state density can be calculated from Eq (\ref{eq-pair-rho-from-ab}) where {\bf a} and {\bf b} can be obtained from Eq (\ref{eq-pair-solutiona+}), Eq (\ref{eq-pair-solutiona-}) and Eq (\ref{eq-pair-solutionb}).

\subsection*{Case 1: No positive feedback ($q=0$)}
The master equation for the singlet and doublet density reduces to the contact process when $q=0$ \cite{matsuda1992PTP}. We know that the contact process model undergoes a continuous phase transition~\cite{marro2005nonequilibrium}, where density of active cells (vegetation density in our case) decreases to zero continuously. Therefore, at the critical point ($p_c$), the density ($\rho_1$) goes to zero. Substituting $q=0$ and $p=p_{c}$ in Eq (\ref{eq-pair-a}) 

$$ -p_c {\bf a} + d = 0 \implies {\bf a} = \frac{d}{p_c} $$

Because $\rho_1 = 0$ at $p=p_{c}$, Eq (\ref{eq-pair-rho-from-ab}) implies ${\bf b} = 0.$ Substituting $q=0$, $p=p_{c}$, $\mathbf{a}=d/p_{c}$ and $\mathbf{b}=0$ in Eq (\ref{eq-pair-solutionb}), assuming $d \neq 0$ and simplifying we get

\begin{equation}
    p_{c} = \frac{4}{3}d
    \label{eq-pair-p-c-contact-process}
\end{equation}

Thus, given that the contact process exhibits a continuous phase transition, the pair approximation predicts a critical point $p_c=\frac{4d}{3}.$ This result is consistent with the results of ~\cite{matsuda1992PTP}. 

\subsection*{Case 2: With positive feedback ($q>0$)}
Now, we investigate the role of positive feedback on phase transition in our model. For $q>0$, $\mathbf a$ will have two solutions given by Eq (\ref{eq-pair-solutiona+}) and Eq (\ref{eq-pair-solutiona-}). Therefore, substituting these in Eq (\ref{eq-pair-solutionb}) to obtain $\mathbf{b}$ and then substituting $\mathbf{a}$ and $\mathbf{b}$ in Eq (\ref{eq-pair-rho-from-ab}), vegetation density will have two non-zero values for some values of $p$ and $d$. From mean-field approximation, we know that one of these solutions is stable and the other is unstable. The region in parameter space where these two solutions (one stable and one unstable) coexist and are positive will show the discontinuous transition. Indeed for a fixed $d=0.3$, the system shows continuous transition for low values of $q$ and discontinuous transition for high values of $q$.

We define the critical point $p=p_{c}$ as the point where vegetation density drops to zero. At the critical point for discontinuous transition,  two non-zero solutions of $\rho_1$ (one stable and one unstable) meet. This occurs where determinant in Eq (\ref{eq-pair-solutiona+}) or Eq (\ref{eq-pair-solutiona-}) vanishes. 

$$ \left[p_c +q(1-d)\right]^2 - 4q(1-q)d  = 0   $$

Since $p_c, q$ and $d$ are non-negative, we have 
$$ p_c = \sqrt{4q(1-q)d} - (1-d)q $$
 
\begin{figure}
\resizebox{1\columnwidth}{!}{%
 \includegraphics{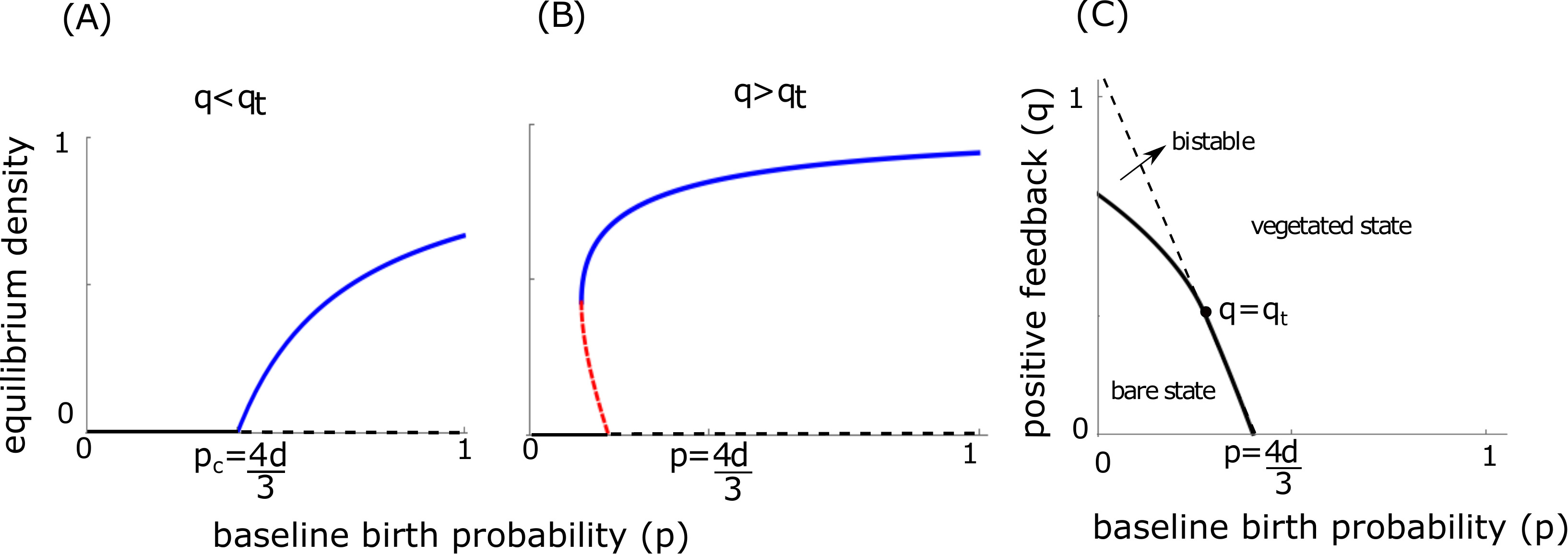} }
\caption{\textbf{Bifurcation diagram obtained by pair approximation analysis.} (A) and (B) show the continuous and discontinuous transitions at $q<q_t$ and $q>q_t$ respectively. (C) is the full phase diagram as a function of $q$ and $p$.The black dot represents the tri-critical point $q=q_t$ at which the nature of transition changes from continuous to discontinuous. The region between solid and dotted black lines is the bistable region }. 
\label{fig-pair-approx}       %
\end{figure}

At the tri-critical point ($q=q_t$), the transition changes from continuous to discontinuous (see \ref{fig-pair-approx}). Therefore, at this point, the critical density (defined as the density at which transition occurs) is zero. Substituting these values in Eq (\ref{eq-pair-solutionb}) for $d=0.3$, we get, $ (p_t,q_t)=(0.27,0.36) $. 

 In the continuous transition regime ($q<q_t$), unlike the mean-field approximation, critical point decreases as a function of $q$. This shows that in our model, positive feedback in the systems with local spatial interactions helps the system sustain its vegetated state in harsh conditions which are represented as low values of $p$. Note that we did not perform stability analysis for this model. However, it is reasonable to assume that the stability of the equilibria will remain the same as the mean-field model. The bifurcation diagram obtained by the pair approximation is shown in Fig \ref{fig-pair-approx}. It is qualitatively same as the output of mean-field model. However, it is clear that the vegetated state is sustained for harsher conditions because there is a reduction in the area of the bare state region. The region of bistability is also reduced in this approximation. Therefore, it can be concluded that local spatial interactions increase the resilience of the system as compared to well-mixed system. This effect of local spatial interactions is the opposite of the effect of the demographic noise as shown in Fig~\ref{fig-finite-analytical}. The real system with both the local interactions and finite-size can show the dynamics resulting from the interplay between these two effects.    
 
\end{appendices}

\end{document}